\documentclass{PoS}

\newcommand\nvec[1]{\textbf{\emph{#1}}}

\title{On multiple gluon exchange in $J/\psi$ hadroproduction}

\ShortTitle{On multiple gluon exchange...}

\author{\speaker{Leszek Motyka}\thanks{Support of the Polish National Science Centre grant no.\ DEC-2011/01/B/ST2/03643 is gratefully acknowledged.}\\
 Institute of Physics, Jagiellonian University, Reymonta 4, 30-059 Krak\'{o}w, Poland \\
        E-mail: \email{leszek.motyka@uj.edu.pl}}

\author{Mariusz Sadzikowski\\
        Institute of Physics, Jagiellonian University, Reymonta 4, 30-059 Krak\'{o}w, Poland \\
        E-mail: \email{mariusz.sadzikowski@uj.edu.pl}}

\abstract{
We consider a contribution to $J/\psi$ hadroproduction in which the meson production is mediated by three-gluon partonic state, with two gluons coming from the target and one gluon from the projectile. This mechanism involves double gluon density and thus it enters at a non-leading twist, but it is enhanced at large energies due to large double gluon density at small $x$. The three-gluon contribution to $J/\psi$ hadroproduction is calculated within perturbative QCD in the $k_T$ factorization framework, and it is found to provide a significant correction to the standard leading twist cross-section at the energies of the Tevatron or the LHC. The results are given as differential $p_T$-dependent cross-sections for $J/\psi$ polarization components.}

\FullConference{XXII. International Workshop on Deep-Inelastic Scattering and Related Subjects,\\
		28 April - 2 May 2014\\
		Warsaw, Poland}

\begin{document}

\section{Introduction}

Understanding of charmonia and bottomonia production in high energy $pp$ and $p\bar p$ collisions has been a challenge for the theory of strong interactions since first Tevatron measurements were published of prompt $J/\psi$ production \cite{Lansberg}. The Tevatron data showed huge excess of the measured cross-section over the simplest LO QCD predictions, reaching almost two orders of magnitude at large $J/\psi$ transverse momenta \cite{TeV0}.  
This na\"{i}ve picture of the process is given by standard, color singlet mechanism (CSM) in the collinear approximation. In this mechanism it is assumed, that all the vector quarkonium quantum numbers and kinematic properties are generated already at the partonic level. The production process of the heavy quark -- anti quark pair is then driven by gluons, $gg \to Q\bar Q g$, and the final state gluon emission of the gluon is necessary to produce the $Q\bar Q$ pair with the quantum numbers of the vector meson. Due to presence of large scales --- the quark mass and large transverse momentum --- the process may be computed within perturbative QCD. In the collinear approximation, the emitted gluon recoil generates the transverse momentum of the meson.  As said above, however, this simple description fails badly to describe the data and needed to be modified. In particular, the transverse momentum dependence of the collinear CSM is much too steep.

\subsection{Color octet model} 

Since it become clear that the collinear color singlet approximation is insufficient, a few mechanisms have been proposed that are able to explain reasonably well the measured heavy quarkonia production. Currently two most successful approaches to prompt vector meson hadroproduction are based on perturbative QCD. The first one is based on the so-called color octet mechanism (COM), that assumes that the $Q\bar Q$ pair may be produced with quantum numbers different from the final state meson, in particular in the color octet representation \cite{Bodwin, CL1}. Originally the COM has been motivated by the heavy quarkonium effective theory in which higher Fock states of the quarkonium wave function may contain an octet $Q\bar Q$ pair with some amplitude of the order of $\alpha_s$. Probably, it may be also viewed as a fragmentation process of partonic $Q\bar Q$ state into the meson, somewhat similar to $c$ or $b$ quark fragmentation into $D$ or $B$ mesons. What is crucial for effective predictivity, in the COM one assumes existence of universal transition amplitudes from the $Q\bar Q$ pair with given quantum numbers to the mesons, provided matching of quark and meson kinematic variables. The COM has been developed up to the next-to-leading order (NLO) within collinear approximation \cite{BK1,BK-NLOfit,BK-pol}, and it was shown to provide a good global fit of prompt quarkonium production data \cite{BK-NLOfit}, except of the quarkonium polarization \cite{BK-pol}. The COM model does not provide a satisfactory description of prompt $J/\psi$ polarization as a function of transverse momentum neither at the leading order nor at the NLO. Therefore, however sound and sophisticated, the COM may be still incomplete.

\subsection{$k_T$ factorisation approach}

The other rather successful approach to vector quarkonia hadroproduction attempts to explain the wide transverse momentum distribution by taking into account the non-zero transverse momentum of initial state gluons assuming the $k_T$-factorization (or high energy factorization) \cite{HKSST, Baranov1,Baranov2,Baranov3,KVS}. The most interesting quarkonia production at the Tevatron and the LHC occur in the kinematic domain where the incoming parton (gluon) momentum fractions $x_i$ are small, typically $x_i < 0.01$. In this region radiation in the QCD evolution may generate sizable transverse momentum, $\vec k$, of the gluons, resulting then with a modified (broader) distribution of quarkonium transverse momentum. In the $k_T$-factorization approach the standard, collinear parton densities, e.g.\ $xg(x,\mu^2)$ for the gluon, are replaced by unintegrated parton densities, e.g. $f(x,k^2,\mu^2)$, with a LL constraint $xg(x,\mu^2) = \int dk^2 / k^2 f(x,k^2,\mu^2)$. The $k_T$-factorization approach was applied to prompt quarkonia production at the LO, assuming both color singlet and color octet scenarios. The emerging picture is quite encouraging, but not fully clear yet. A recent description of the recent LHC data within the $k_T$-factorization CSM approach was quite successful, including even meson polarization \cite{Baranov3}, but the older description of the Tevatron data required also the color octet contributions \cite{Baranov1}. Therefore, in this moment the $k_T$-factorizing approach also suffers from some deficiencies in providing the consistent global picture of prompt quarkonia production.

\section{The triple gluon correction}

\subsection{Motivation}

As discussed in the previous section, the existing approaches to vector quarkonia hadroproduction, although reproducing well many features of the data, do not provide complete and fully satisfactory global description. Thus it may be still necessary to include yet other production mechanism. One of the potentially important corrections to the standard CSM and COM contributions may be driven by multiple scattering effects. At the lowest order (assuming matching of quark and meson quantum numbers) the correction is driven by a partonic amplitude of three-gluon fusion into heavy quark-antiquark pair, $g + 2g \to Q \bar Q$ \cite{KMRS}. Since one needs to drag two glouns out of one hadron, this correction employs the double gluon distribution and enters, therefore, beyond the leading twist. It implies a power suppression of this correction w.r.t.\ the standard, two-gluon cross-sections. The double gluon distribution, however, at small gluon $x$, provides a large enhancement factor $\sim xg(x,\mu^2)$, that may well reach about 20 in the relevant kinematic domain. The first estimates suggested that this multiple scattering effect may be even the dominant contribution of the total vector quarkonia cross-section, leaving only little space for the color octet contributions \cite{KMRS}. Moreover, the multiple scattering effects should be also important in vector meson productions in high energy collisions with nuclei, where the nucleus mass number provides an additional enhancement factor of multiple gluon distributions. Therefore we performed a detailed estimate of this rescattering (or triple gluon) correction for the Tevatron and the LHC energies beyond the collinear, leading logarithmic accuracy of Ref.\ \cite{KMRS}.

\subsection{The framework}

\begin{figure}
\centerline{
\begin{tabular}{ll}
a) \hspace{1em}\includegraphics[width=0.32\columnwidth]{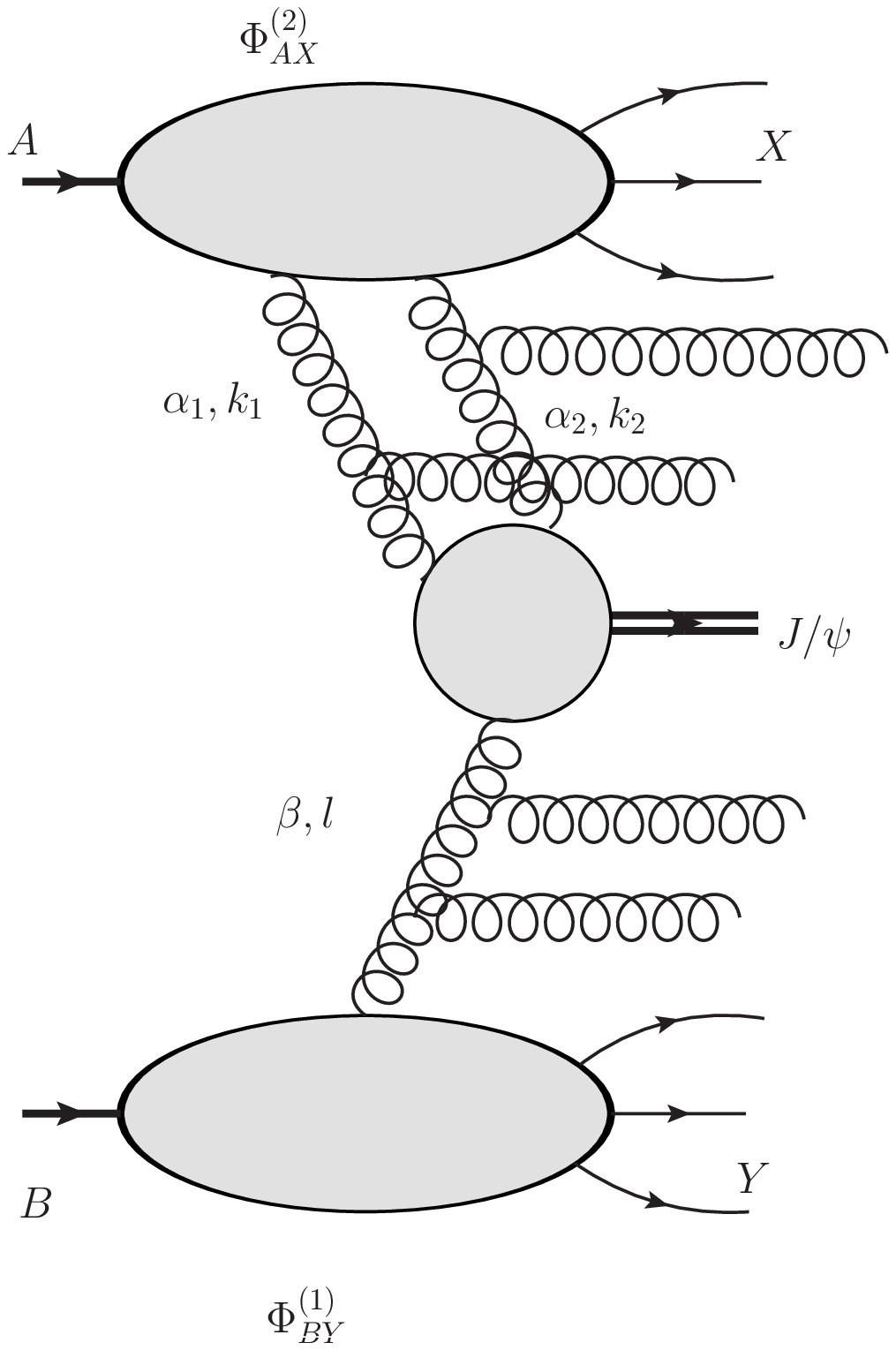}\hspace{1em} & 
b) \hspace{1em}\includegraphics[width=0.48\columnwidth]{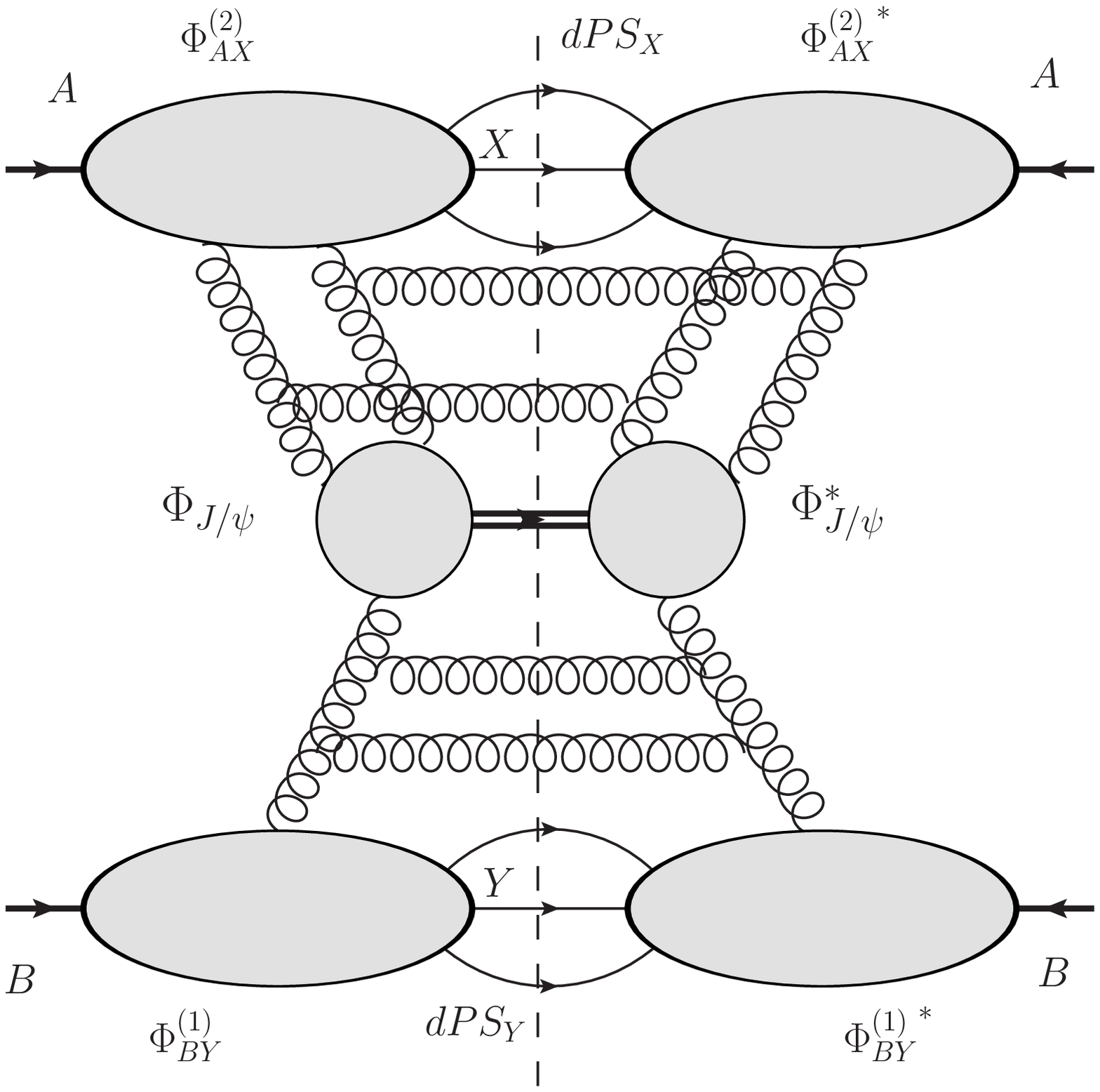} 
\end{tabular}
}
\caption{Diagrams relevant for the triple-gluon mechanism of heavy vector meson hadroproduction: a) the amplitude and b) the dominant topology in the cross-section.  
\label{fig1}
}
\end{figure}

The calculations were performed within the $k_T$-factorization approach. The relevant diagram for an inclusive process, $AB \to XY J/\psi$ is shown in Fig.\ 1a. The total amplitude includes amplitudes of this diagram and the `upside-down' diagram in which the two gluons originate from projectile $B$. When squaring the amplitude, one finds diagrams with topology shown in Fig.\ 1b, with four and two gluons in the $t$-channel, and (not shown) diagrams with three gluon $t$-channel states originating from both $A$ and $B$. The latter `interference' diagrams may be neglected as the three gluon state evolution is known to have no energy enhancement, contrary to the two and four gluon states that drive the contributions shown in Fig.\ 1b. After performing the phase space integrals over remnants $X$ and $Y$ in this contribution one recovers double and single unintegrated gluon distributions originating from $A$ and $B$ respectively. The three gluon-fusion amplitude into the meson (the middle part of the diagram shown in Fig.\ 1a) is described in terms of a known impact factor, $\Phi_{J/\psi}$\cite{Bzdak}, with $J/\psi$ polarization vectors in the helicity frame. The proper normalization of the amplitudes is obtained by matching to collinear cross-sections of single and double parton scattering. We assume factorization of the double gluon density and factorization of the impact parameter dependence. With these assumptions the final formula for the triple gluon contribution to $J/\psi$ production takes the form:
\begin{equation}
\label{cross-section2}
\frac{d^2\sigma_{pp\rightarrow J/\psi X}}{dY dp_T^2} 
= {\cal N} \, {R_{\mbox{sh}}^2 \over \sigma_{\mathrm{eff}}} \, 
\int d^2\nvec{k}d^2\nvec{l}\,
\frac{ \alpha_s^3\, f(\beta,l^2,\mu) \,
f(\alpha,k^2,\mu) \, f(\alpha,(\nvec{p}-\nvec{k}-\nvec{l})^2,\mu)}
{\left[k^2\,l^2\,(\nvec{p}-\nvec{k}-\nvec{l})^2\right]^2} 
\]
\[
\times \, \left|\Phi_{J/\psi}(\alpha,\beta;\nvec{k},\nvec{p}-\nvec{k}-\nvec{l}, \nvec{l};\epsilon)\right|^2
\; + \; (\alpha\leftrightarrow\beta,\; p_A\leftrightarrow p_B) 
\end{equation}
where $\alpha$ and $\beta$ are meson longitudinal momentum fractions of the projectiles, rapidity $Y = 1/2 \log(\alpha/\beta)$, $\nvec{p}$ is the meson transverse momentum, $p_T = |\nvec{p}|$, $\epsilon$ is the meson polarization, $\nvec{k}$, $\nvec{l}$ and $\nvec{p} - \nvec{k} - \nvec{l}$ are the gluons transverse momenta, and we combine numerical constants, color factors etc. into a normalization constant ${\cal N}$.

Note the emergence of $\sigma_{\mathrm{eff}}$, the effective parameter describing parton transverse density and probability of multiple scattering. It appears as a result of forcing two partons from one hadron to be in the same position in the transverse space. Since, in the double gluon density the values of gluon~$x$ are different, we used off-diagonal gluon densities and included the Shuvaev factor, $R_{\mathrm{sh}}$ \cite{Shuvaev}.

\subsection{Results}

\begin{figure}
\centerline{
\begin{tabular}{ll}
a) \hspace{1em}\includegraphics[width=0.47\columnwidth]{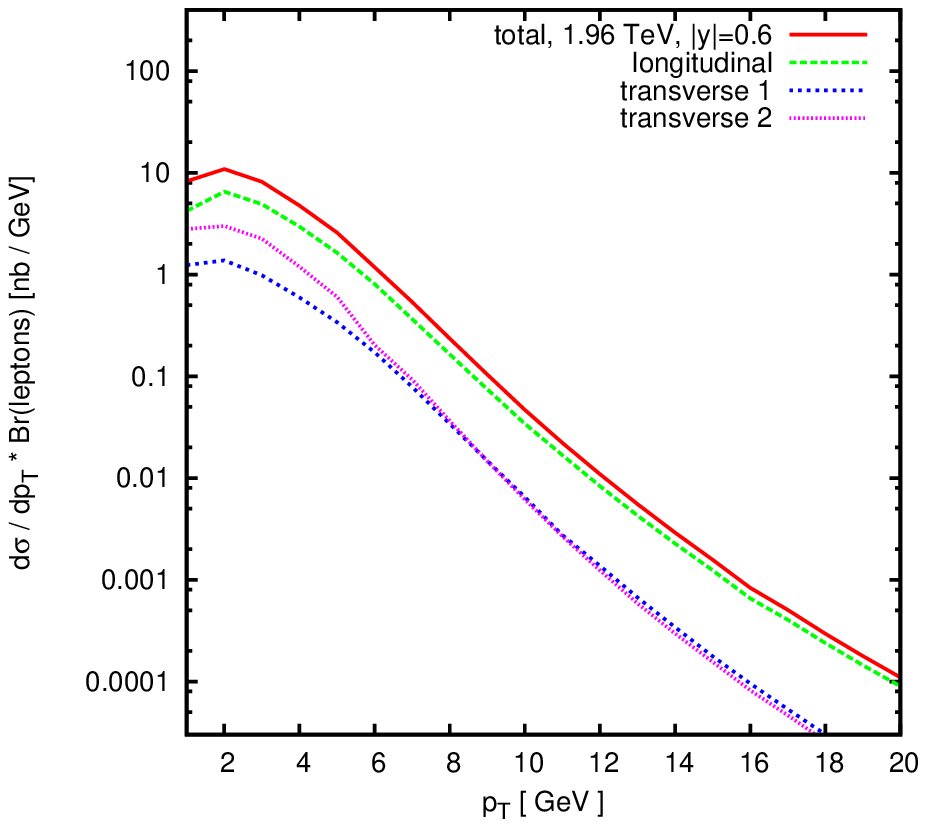}\hspace{1em} & 
b) \hspace{1em}\includegraphics[width=0.47\columnwidth]{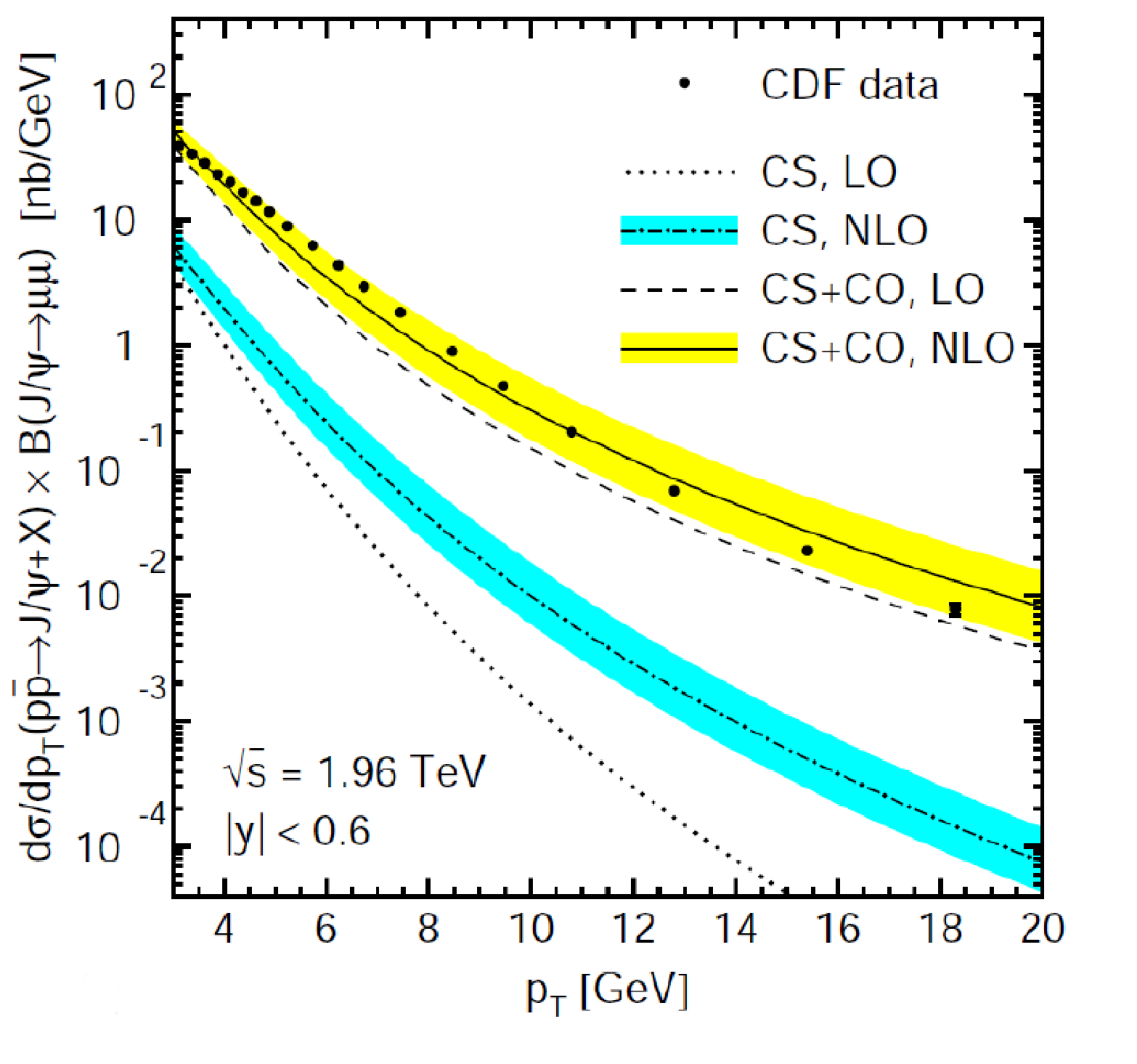} 
\end{tabular}
}
\caption{Differential cross sections $\left. \frac{d\sigma_{pp\rightarrow J/\psi X}}{dp_T}\right|_{|Y|<0.6} \times \mathrm{Br}(J/\psi \to \mu^+ \mu^-)$ at the Tevatron, $\sqrt{s} = 1.96$~TeV: a) the triple gluon correction (this paper) and its break-down into polarized components b) the reference: plot taken from arXiv:1105.0820 [hep-ph]\cite{BK-NLOfit}: CDF data \cite{CDFdata} and LO / NLO fits of the CSM and COM \cite{BK-NLOfit}. 
\label{fig2}
}
\end{figure}

In the numerical evaluations unintegrated gluon densities were used derived from the CT10 collinear gluon density \cite{ct10} using Kimber-Martin-Ryskin approach~\cite{KMR} with the hard scale given by the transverse meson mass, $\mu^2 = M_{J/\psi} ^2 + p_T^2$. The running strong coupling constants of a gluon with virtuality $k^2$ was evaluated at the scale $\mu^2 = M_c^2 + k^2$, with $M_c = M_{J/\psi} /2$. We set the multiple scattering parameter value $\sigma_{\mathrm{eff}} = 15$~mb, in accordance with Refs. \cite{sigma0cdf,sigma0d0}.

In Fig.\ 2a the results of numerical evaluation of the triple gluon correction to the differential meson production cross section 
$\left. \frac{d\sigma_{pp\rightarrow J/\psi X}}{dp_T}\right|_{|Y|<0.6} \times \mathrm{Br}(J/\psi \to \mu^+ \mu^-) $ (including the branching ratio of the meson decay to muons) are shown for the Tevatron energy ($\sqrt{s} = 1.96$ GeV). For a reference, we show in Fig.\ 2b a plot taken from Ref. \cite{BK-NLOfit}, in which results of CSM and of COM fits are presented at LO and NLO accuracy together with CDF data. From the comparison of Fig.\ 2a and Fig.\ 2b it is clear, that the triple gluon correction is similar in the magnitude and shape to the standard CSM contribution at the NLO. In more detail, the triple-gluon contribution is found to be larger than the CSM contribution in the whole range of transverse momentum and it reaches about 20 -- 25\% of the measured cross-section at low transverse momenta. At larger transverse momenta this contributions the relative importance of this correction diminishes and already at $p_T > 10$~GeV, this contribution is negligible.

In Fig.\ 2a we also show polarized components of the triple gluon correction in the helicity frame. Clearly, at low transverse momentum the longitudinal and total transverse cross-sections are similar, and with increasing $p_T$ the longitudinal component becomes dominant, saturating the total cross-section.

The results for the LHC exhibit very similar pattern, as the results for the Tevatron, so we do not show them here, leaving the broader and more detailed presentation to the forthcoming paper \cite{future}.

\section{Conclusions}

We estimated corrections to prompt $J/\psi$ production at the Tevatron and the LHC coming from a triple gluon incoming partonic state. They are found to be sizable at small transverse momenta, up to 20 -- 25 \% of the measured cross-section and negligible at large transverse momenta. The $J/\psi$ polarization from this mechanism varies with $p_T$, from negligible polarization at small momenta to longitudinal polarization at large $p_T$. The correction is large enough to affect the fits to the existing, precise data on $J/\psi$ hadroproduction. The results may be also relevant for $J/\psi$ production in proton--nucleus collisions, where the multiple gluon exchanges are enhanced by the nucleus mass number.

\end{document}